\documentclass[twocolumn]{aastex701}

\newcommand{\clname}{SPT-CL\,J0546-5345}
\newcommand{\Msol}{\ensuremath{M_{\odot}}}

\shorttitle{A Massive strong-lensing cluster at $z = 1.07$}
\shortauthors{J. Allingham et al.}

\usepackage{amsmath}

\tabletypesize{\footnotesize}
\begin{document}

%%%%%%%%%%%%%%%%%%%%%%%%%%%%%%%%%%%%%%%%%%%%%%%%%%%%
\title{SLICE: \clname\ -- A prominent strong-lensing cluster at $z=1.07$}

%%%%%%%%%%%%%%%%%%%%%%%%%%%%%%%%%%%%%%%%%%%%%%%%%%%%
\correspondingauthor{Joseph Allingham}
\email{allingha@post.bgu.ac.il}

\author[0000-0003-2718-8640]{Joseph F. V. Allingham}
\affiliation{Department of Physics, Ben-Gurion University of the Negev, P.O. Box 653, Be'er-Sheva 84105, Israel}
\email{allingha@post.bgu.ac.il}

\author[0000-0002-0350-4488]{Adi Zitrin}
\affiliation{Department of Physics, Ben-Gurion University of the Negev, P.O. Box 653, Be'er-Sheva 84105, Israel}
\email{zitrin@bgu.ac.il}

\author[0000-0001-9411-3484]{Miriam Golubchik}
\affiliation{Department of Physics, Ben-Gurion University of the Negev, P.O. Box 653, Be'er-Sheva 84105, Israel}
\email{golubmir@post.bgu.ac.il}

\author[0000-0001-6278-032X]{Lukas J. Furtak}
\affiliation{Department of Physics, Ben-Gurion University of the Negev, P.O. Box 653, Be'er-Sheva 84105, Israel}
\email{furtak@post.bgu.ac.il}

\author[0000-0003-1074-4807]{Matthew Bayliss}
\affiliation{Department of Physics, University of Cincinnati, Cincinnati, OH 45221, USA}
\email{matthew.bayliss@uc.edu}

\author[0000-0002-8261-9098]{Catherine Cerny}
\affiliation{Department of Astronomy, University of Michigan 1085 South University Avenue Ann Arbor, MI 48109, USA}
\email{cecerny@umich.edu}

\author[0000-0001-9065-3926]{Jose M. Diego}
\affiliation{Instituto de F\'isica de Cantabria (CSIC-UC). Avda. Los Castros s/n. 39005 Santander, Spain}
\email{jdiego@ifca.unican.es}

\author[0000-0002-3398-6916]{Alastair C. Edge}
\affiliation{Centre for Extragalactic Astronomy, Department of Physics, Durham University, South Road, Durham DH1 3LE, UK}
\affiliation{Institute for Computational Cosmology, Durham University, South Road, Durham DH1 3LE, UK}
\email{alastair.edge@durham.ac.uk}

\author[0009-0004-7337-7674]{Raven Gassis}
\affiliation{Department of Physics, University of Cincinnati, Cincinnati, OH 45221, USA}
\email{gassismr@mail.uc.edu}

\author[0000-0003-1370-5010]{Michael D. Gladders}
\affiliation{Kavli Institute for Cosmological Physics, University of Chicago, 5640 South Ellis Avenue, Chicago, IL 60637, USA}
\affiliation{Department of Astronomy and Astrophysics, University of Chicago, 5640 South Ellis Avenue, Chicago, IL 60637, USA}
\email{gladders@uchicago.edu}

\author[0000-0003-1974-8732]{Mathilde Jauzac}
\affiliation{Centre for Extragalactic Astronomy, Department of Physics, Durham University, South Road, Durham DH1 3LE, UK}
\affiliation{Institute for Computational Cosmology, Durham University, South Road, Durham DH1 3LE, UK}
\affiliation{Astrophysics Research Centre, University of KwaZulu-Natal, Westville Campus, Durban 4041, South Africa}
\affiliation{School of Mathematics, Statistics \& Computer Science, University of KwaZulu-Natal, Westville Campus, Durban 4041, South Africa}
\email{mathilde.jauzac@durham.ac.uk}

\author[0000-0002-7633-2883]{David J. Lagattuta}
\affiliation{Centre for Astrophysics Research, Department of Physics, Astronomy and Mathematics, University of Hertfordshire, Hatfield AL10 9AB, UK}
\affiliation{Centre for Extragalactic Astronomy, Department of Physics, Durham University, South Road, Durham DH1 3LE, UK}
\affiliation{Institute for Computational Cosmology, Durham University, South Road, Durham DH1 3LE, UK}
\email{david.j.lagattuta@durham.ac.uk}

\author[0009-0004-2523-4425]{Gavin Leroy}
\affiliation{Institute for Computational Cosmology, Durham University, South Road, Durham DH1 3LE, UK}
\email{gavin.leroy@durham.ac.uk}

\author[0000-0001-6636-4999]{Marceau Limousin}
\affiliation{Aix Marseille Univ, CNRS, CNES, LAM, Marseille, France}
\email{marceau.limousin@lam.fr}

\author[0000-0003-3266-2001]{Guillaume Mahler}
\affiliation{STAR Institute, Quartier Agora - All\'ee du six Ao\^ut, 19c B-4000 Li\`ege, Belgium}
\email{Guillaume.Mahler@uliege.be}

\author[0000-0002-7876-4321]{Ashish K. Meena}
\affiliation{Department of Physics, Ben-Gurion University of the Negev, P.O. Box 653, Be'er-Sheva 84105, Israel}
\email{meena@post.bgu.ac.il}

\author[0000-0002-5554-8896]{Priyamvada Natarajan}
\affiliation{Department of Astronomy, Yale University, New Haven, CT 06511, USA}
\affiliation{Department of Physics, Yale University, New Haven, CT 06511, USA}
\email{priyamvada.natarajan@yale.edu}

\author[0000-0002-7559-0864]{Keren Sharon}
\affiliation{Department of Astronomy, University of Michigan 1085 South University Avenue Ann Arbor, MI 48109, USA}
\email{kerens@umich.edu}

%%%%%%%%%%%%%%%%%%%%%%%%%%%%%%%%%%%%%%%%%%%%%%%%%%%%
\begin{abstract}
Massive galaxy clusters act as prominent strong-lenses. Due to a combination of observational biases, cluster evolution and lensing efficiency, most of the known cluster lenses lie typically at $z_{l}\sim0.2-0.7$, with only a few prominent examples at higher redshifts.
Here we report a first strong-lensing analysis of the massive galaxy cluster \clname\ at a redshift $z_l=1.07$.
This cluster was first detected through the Sunyaev-Zel'dovich effect, with a high estimated mass for its redshift of $M_{200,c} = (7.95 \pm 0.92) \times 10^{14}\,\Msol$.
Using recent JWST/NIRCam and archival HST imaging, we identify at least 10 secure and 6 candidate sets of multiply imaged background galaxies, which we use to constrain the mass distribution in the cluster.
We derive effective Einstein radii of $\theta_{E}= 18.1 \pm 1.8 \arcsec$ for a source at $z_{s}=3$, and $\theta_{E}= 27.9 \pm 2.8 \arcsec$ for a source at $z_{s}=9$.
The total projected mass within a $200$\,kpc radius around the strong-lensing region is $M(<200\,\mathrm{kpc}) = (1.9 \pm 0.3) \times 10^{14}\,\Msol$. 
While our results rely on photometric redshifts warranting spectroscopic follow-up, this central mass resembles that of the Hubble Frontier Fields clusters -- although \clname\ is observed when the Universe was $\sim 3-4$\,Gyr younger.
Amongst the multiply-imaged sources, we identify a hyperbolic-umbilic-like configuration, and, thanks to its point-like morphology, a possible Active Galactic Nucleus (AGN). If confirmed spectroscopically, it will add to just a handful of other quasars and AGN known to be multiply lensed by galaxy clusters. 
\end{abstract}

%%%%%%%%%%%%%%%%%%%%%%%%%%%%%%%%%%%%%%%%%%%%%%%%%%%%

%%%%%%%%%%%%%%%%%%%%%%%%%%%%%%%%%%%%%%%%%%%%%%%%%%%%
\section{Introduction}\label{sec:intro}
Structure in the Universe is believed to grow hierarchically, with smaller structures forming first, later coalescing to form larger structures \citep[e.g.][]{1982Natur.300..407Z,Springel2005Natur.435..629S}. As the largest and most massive gravitationally bound objects in the Universe, galaxy clusters form late in cosmic history, around $z\sim3$, when the Universe was about 2 Gyr old \citep[e.g.][]{2012ARA&A..50..353K, 2015SSRv..188...93P}.

Dedicated imaging campaigns with the Hubble Space Telescope targeting massive galaxy clusters have revealed that they act as prolific gravitational lenses (e.g. CLASH, \citealt[][]{PostmanCLASHoverview, Zitrin2014CLASH25}; Hubble Frontier Fields, \citealt[][]{Lotz2016HFF}; RELICS, \citealt[][]{Coe2019RELICS}; BUFFALO, \citealt{Steinhardt2020BUFFALO}). The lensing efficiency of clusters depends on various factors such as their redshift, their mass, and its distribution. Lensing is dictated by the projected mass density such that more evolved clusters that are typically more concentrated, as well as those elongated along the line-of-sight, tend to show more prominent lensing features and exhibit large critical areas \citep{Zitrin2013M0416}. It has also been shown that massive, merging clusters tend to show accentuated lensing properties, owing to the various massive substructures that together form a larger lens \citep[e.g.,][]{Meneghetti2003,Zitrin2017,Mahler2019MACS0417, 2019MNRAS.483.3082J,Acebron2019RXC0032,Furtak2022UNCOVER}.

Given the above, most well-known lensing clusters typically lie at redshifts $z_{l}\gtrsim0.2$, below which the lensing efficiency becomes substantially weaker as the extended lens gets too close to the observer, and $z_{l}\lesssim0.7$. This upper empirical ``limit'' is a result of various factors including cluster evolution, the change in lensing efficiency with redshift, and the typical observational depths. 
Also, a source sitting closely behind a $z_{l}\sim1$ cluster would not be efficiently lensed, but may be prominently lensed by lower redshift clusters (see e.g.\ the giant arc in Abell 370, \citealt{Richard2010A370}), making these lenses easier to identify.
Deeper or longer-wavelength surveys, such as those enabled by JWST, are needed to detect the bulk of high-redshift galaxies that are possibly lensed by high-redshift clusters.

In addition, massive lensing clusters are often optically, near infrared \cite[e.g.][]{Gladders2005RCS,Rykoff2014Mapper} or X-ray selected \citep[e.g.][]{Ebeling2010FinalMACS}, which also limits their redshift range.
In that respect, Sunyaev-Zel'dovich effect (SZE) surveys have the potential to detect higher redshift clusters given the (relative) insensitiveness of SZE to redshift. 
Indeed, surveys with the Planck spacecraft \citep{Planck2011highzClus}, South Pole Telescope \citep[SPT,][]{BleemSPTCat2015ApJS..216...27B}, and Atacama Cosmology Telescope \citep[ACT,][]{HiltonACTCat2021ApJS..253....3H} have detected numerous such candidates.

The massive galaxy cluster \clname\ was first discovered through the SZE with the SPT \citep{2009ApJ...701...32S, SPT2010ClustersV1}. \citet{Brodwin2010ApJ...721...90B} measured a cluster redshift of $\langle z_l \rangle = 1.067$, making \clname\ the first $z > 1$ cluster discovered by the SZE. Combining velocity dispersion with X-ray, SZE, and richness measures, they derived a high mass approaching $M_{200,c}\sim1\times 10^{15}\,\Msol$ (see also \citealt{2025arXiv250512110A}). While some lensing features were noted in imaging data as early as \citet{2009ApJ...701...32S}, \clname\ has never been, to our knowledge, subject to strong lens modeling.

Here, we present a first strong-lensing analysis of the cluster, now enabled thanks to new JWST observations from the Strong LensIng and Cluster Evolution program (SLICE; JWST Cycle 3 GO-05594, PI: Mahler; G. Mahler et al. in prep.; \citealt{Cerny25}) along with archival HST and ground-based data. 
Our work adds to a recent analysis which included four $z \in [0.8; 1.06]$ clusters presented in \cite{Cerny25}, and several other strong lensing galaxy clusters around $z_l \sim 1$ analyzed in the past years \citep[see e.g.][]{Zitrin2014CLASH25, 2018ApJ...863..154P, 2020ApJ...894..150M, Diego2022Quyllur, smith2025z103mergingclustersptcl}.

This work is organized as follows: in \S \ref{sec:data} we describe the data used in this work. In \S \ref{sec:analysis} and \ref{sec:discussion} we describe the analysis of the cluster and discuss the results, respectively. 
Our conclusion is presented in \S \ref{sec:conclusion}. Throughout this Letter, we use a standard flat $\Lambda$CDM cosmology with $H_0=70~\mathrm{km}\,{\mathrm{s}^{-1}\,\mathrm{Mpc}^{-1}}$, $\Omega_{\Lambda}=0.7$, and $\Omega_\mathrm{m}=0.3$. With these parameters, 1\,arcsecond at the redshift of the cluster corresponds to 8.12\,kpc. Magnitudes are quoted in the AB system \citep{Oke1983ABandStandards}, and all quoted uncertainties represent $1\sigma$ ranges unless stated otherwise.

\begin{figure*}
    \centering
    \includegraphics[width=1\textwidth]{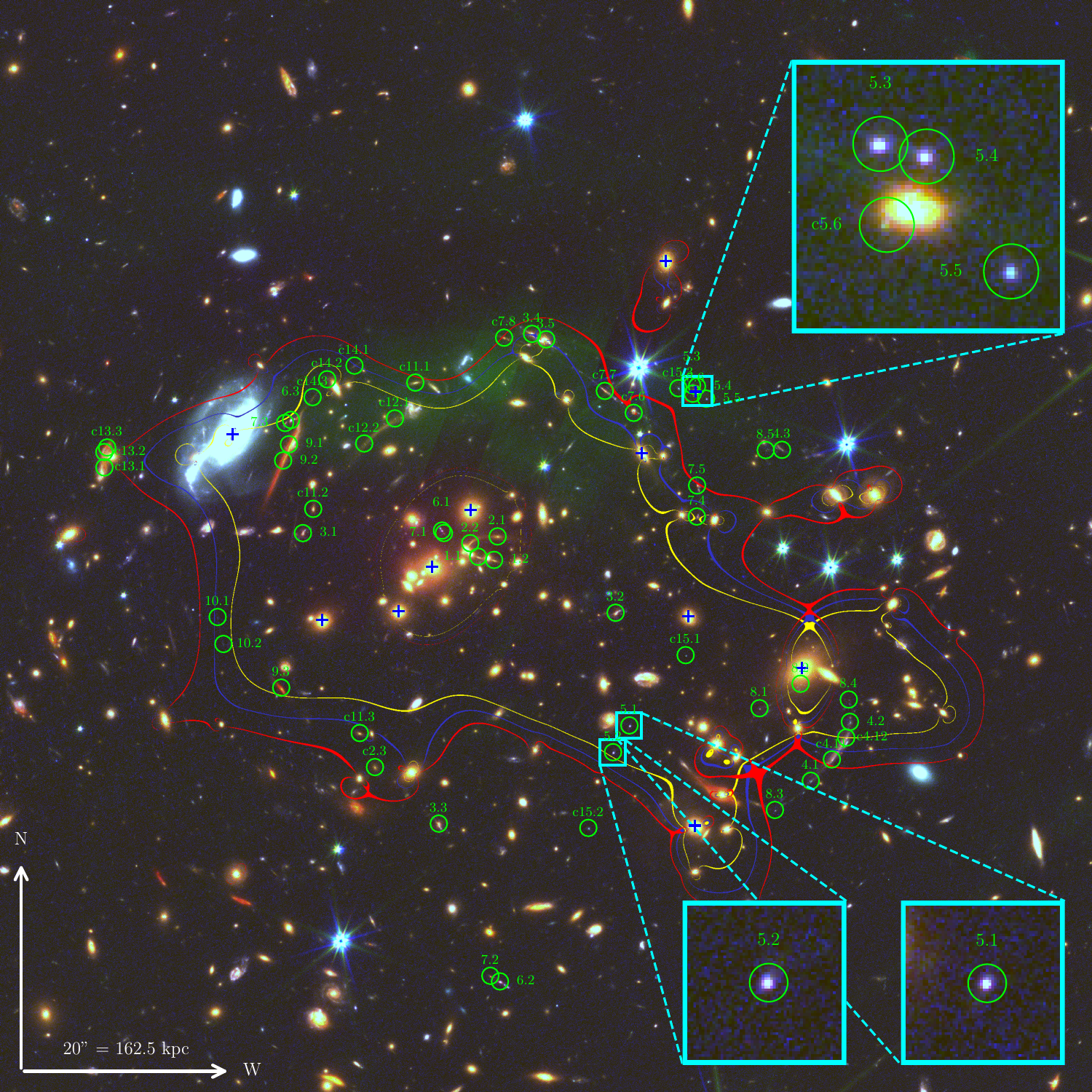}
    \caption{Color-composite image of \clname\ constructed from SLICE JWST and archival HST imaging of the cluster (Red: F322W2; Green: F150W2; Blue: F814W).
    Strong-lensing multiple images are numbered and labeled in green. 
    The yellow, blue and red lines represent, respectively, the critical curves for source redshifts $z_{\mathrm{s}}=3.5$ (System\,3), $z_{\mathrm{s}}=4.5$ and $z_{\mathrm{s}}=6$, as computed from our strong lensing model of the cluster. Galaxies left freely weighted in the modeling are marked with crosses. Highlighted in cyan squares are the images of System\,5, the multiply imaged point-like nucleus -- and potential AGN -- reported in this work.  Note that an artifact from the F150W2 band appears in green north of the cluster core. This is cosmetic only and we verified that it has no substantial effect on our results (see Section \ref{subsec:Photo-z}).}
    \label{fig:cc}
\end{figure*}

\section{Data}\label{sec:data}

\subsection{Imaging and spectroscopy} \label{subsec:imaging_and_spectro}

\clname\ was observed with JWST in Cycle 3 (Program ID: GO-5594; PI: G. Mahler) with the Near Infrared Camera \citep[NIRCam;][]{rieke23}. These observations comprise 1836\,s of observing time in both F150W2 and F322W2, taken simultaneously. 
While SLICE performs in-house dedicated data reduction (G. Mahler et al. in prep.; \citealt{Cerny25}), for the current work, we use the data reduced using the Grism redshift \& line analysis software for space-based slitless spectroscopy \citep[\texttt{grizli},][]{Grizli}\footnote{\url{https://github.com/gbrammer/grizli}}
as downloaded from the DAWN JWST Archive (DJA)\footnote{\url{https://dawn-cph.github.io/dja/imaging/v7/}}.
These observations are complemented with HST observations from WFC3 (Cycle 25, Program ID: GO-15294; PI: G. Wilson), amounting to a total of 5623.5\,s integration with the F160W broad-band filter, and from ACS (Cycle19, Program ID: GO-12477; PI: F. High), for a total of 1920\,s integration with the F606W and 1916\,s with the F814W broad-band filters, respectively. The F814W and F160W data are retrieved though the DJA together with the JWST data. The F606W data were not included in the DJA and we retrieve them from the \texttt{MAST} archive directly and align them onto the same pixel grid as the JWST (and other HST) data using the software \texttt{SWarp} \citep{Bertin2002TERAPIX}.
There also exists VLT/VIMOS imaging in the U, B, R, I bands (program 097.A-0734, PI: R. Demarco), for $\sim1500$ -- $4000$\,s in each band, of which we make use here to further aid or cast confidence, mostly on a by-eye basis, in the multiple-image and cluster-member selections.

Spectroscopic measurements are reported in the literature for this cluster, although none for the multiply imaged sources we identify here.
\citet{Brodwin2010ApJ...721...90B} first confirmed spectroscopically 18 early-type cluster members with the IMACS/GISMO instrument on the Magellan Baade telescope \citep{2011PASP..123..288D}. 
\citet{2014ApJ...792...45R, 2016MNRAS.461..248S, Sweet2017MNRAS.464.2910S, Schrabback2018MNRAS.474.2635S, 2021MNRAS.500..358B} continued the spectroscopic analysis of the cluster with observations from e.g.\ GMOS, FORS2, IMACS, LDSS3 and SALT/RSS, amounting to a total of $\sim 100$ spectroscopic cluster member detections. We use a compilation of these spectroscopic redshifts to validate our cluster members selection (Section \ref{subsec:cluster_member_galaxies}).

Some other relevant data exist for the cluster, although not explicitly used here. For example, \cite{2018ApJ...855...26A, 2018ApJ...853..195W} identified a population of dusty, star-forming galaxies in the cluster background, using a combination of ALMA, APEX/LABOCA, ACTA, and the PACS and Spire instruments of the Herschel Space Telescope. 
At last, \citet{2025arXiv250608853P} concluded from MeerKAT observations coupled with Chandra that \clname\ exhibits a mostly relaxed morphology.

\subsection{Cluster member galaxies} \label{subsec:cluster_member_galaxies}

We build a photometric catalog for the cluster using the software \texttt{SExtractor} \citep{BertinArnouts1996Sextractor} from all HST and JWST filters available. As these are the starting point of the lens model (see Section \ref{sec:analysis}), we aim to extract a catalog of cluster member galaxies. To that end we follow the red-sequence method \citep{GladdersYee2000Finder}. We construct a color-magnitude diagram using the ACS/F814W and WFC3/F160W bands (at the cluster's redshift, the break around rest-frame $\sim$4000\AA\ falls in the middle of the F814W filter).
After fitting the main sequence of galaxies down to 23 AB magnitude, we choose sources within 0.5 mag from the fitted line. 
We then check the validity of the selection by comparing to the list of spectroscopically confirmed cluster members (see Section \ref{subsec:imaging_and_spectro}). 
We further refine the catalog using a by-eye inspection of the image to detect spurious or missing galaxies that appear to be cluster members. Our final selection includes 160 cluster member galaxies. We take the F814W magnitude as reference in the associated lens modeling mass-luminosity scaling relation.

\subsection{Photometric redshift measurements} \label{subsec:Photo-z}
Photometric redshifts, especially in the absence of spectroscopic redshifts for most background sources, can help in the identification of multiply imaged systems, and are needed to estimate the sources' redshifts for the lens modeling. 
Here we use the SED fitting software \texttt{Bagpipes} \citep[see][]{Bagpipes2018}, modified with custom stellar libraries from \citet{BC03}, to obtain photometric redshifts for multiple image candidates. We adopt a constant SFH, and a lower limit on redshift of $z=1.2$.
We note that there are only five wide imaging bands, which renders these photometric redshift measurements uncertain, and emphasizes the need for a spectroscopic follow-up of the cluster.

In Figure \ref{fig:cc} we show a color-composite image of the cluster, where there appears to be a green artifact from the NIRCam/F150W2 band. The effect of this artifact on the photometric redshift measurement should be negligible, given the local background subtraction. To test this explicitly, we rerun the photometric-redshift estimation while introducing a baseline $20\%$ error on NIRCam/F150W2 measurements in the relevant region. This yields photometric redshifts well within $1 \sigma$ of results reported in Table\,\ref{tab:multTable}.

\begin{figure*}
    \centering
    \includegraphics[width=0.48\textwidth]{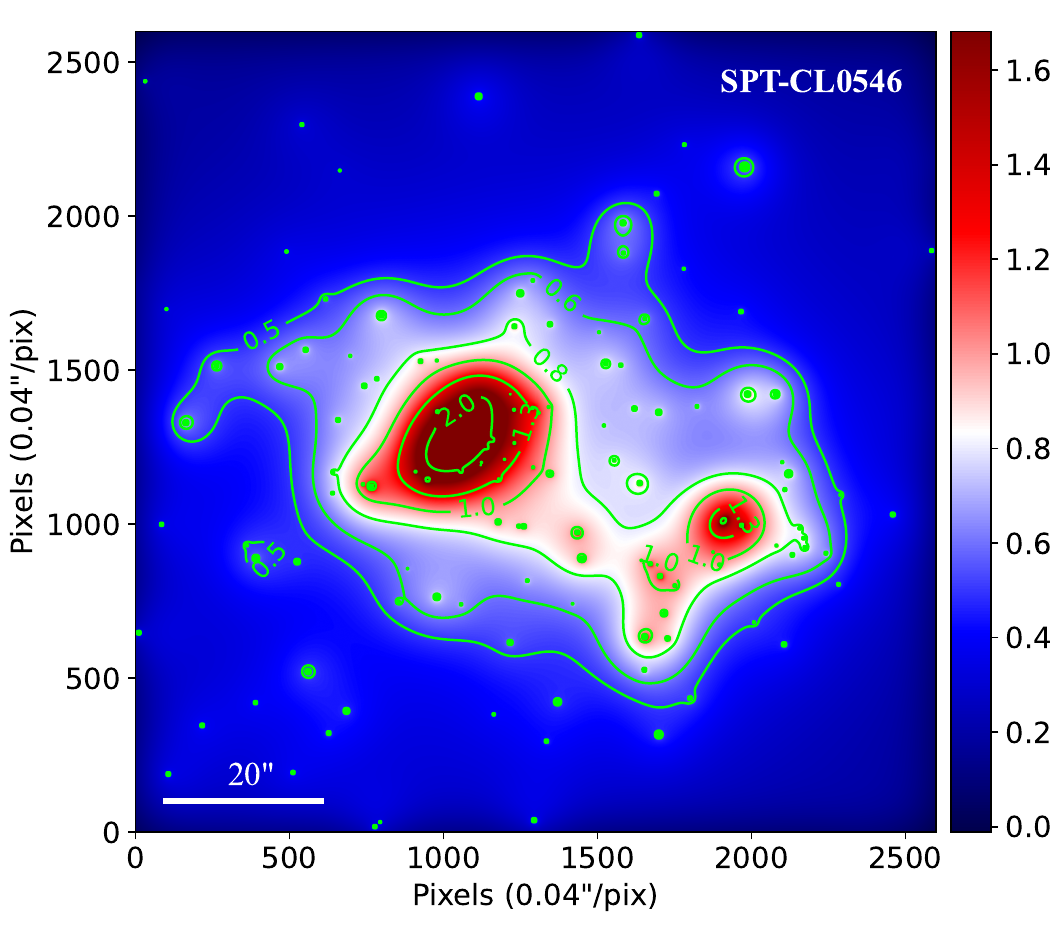}
    \includegraphics[width=0.48\textwidth]{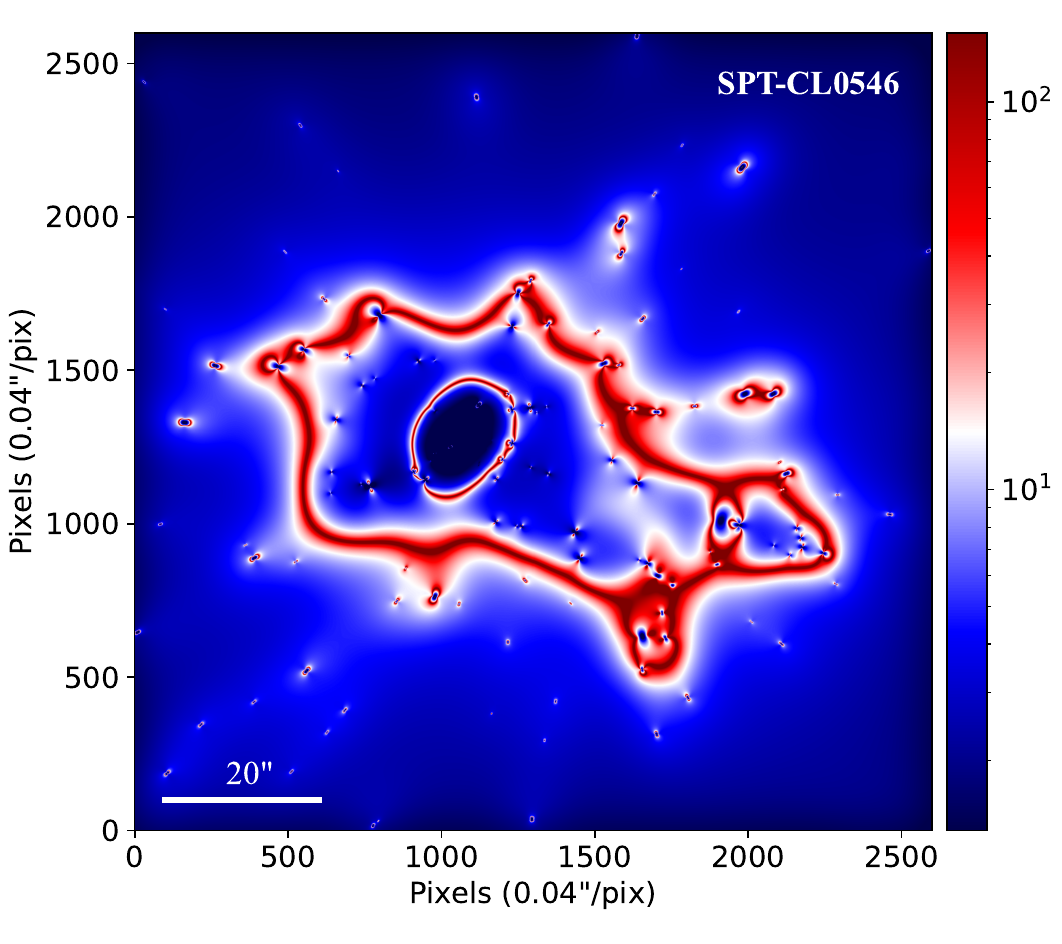}
    \caption{Surface mass density map (left), in units of the critical density for lensing, and magnification map (right), both for an assumed source redshift of $z_s = 3.5$, from the best-fit LTM model.}
    \label{fig:kappamu}
\end{figure*}

\begin{figure*}
    \centering
    \begin{minipage}[t]{0.492\textwidth}
    \vspace{0pt}
        \centering
        \includegraphics[width=\linewidth]{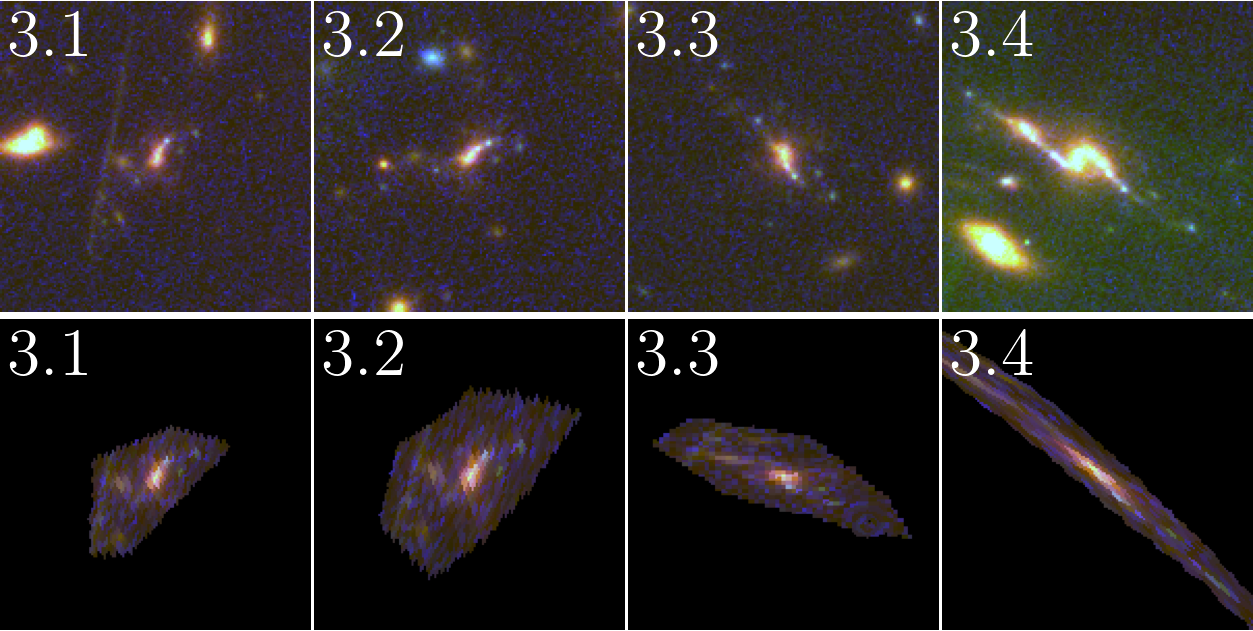}\label{subfig:Sys3}\\[.5em]
        \includegraphics[width=\linewidth]{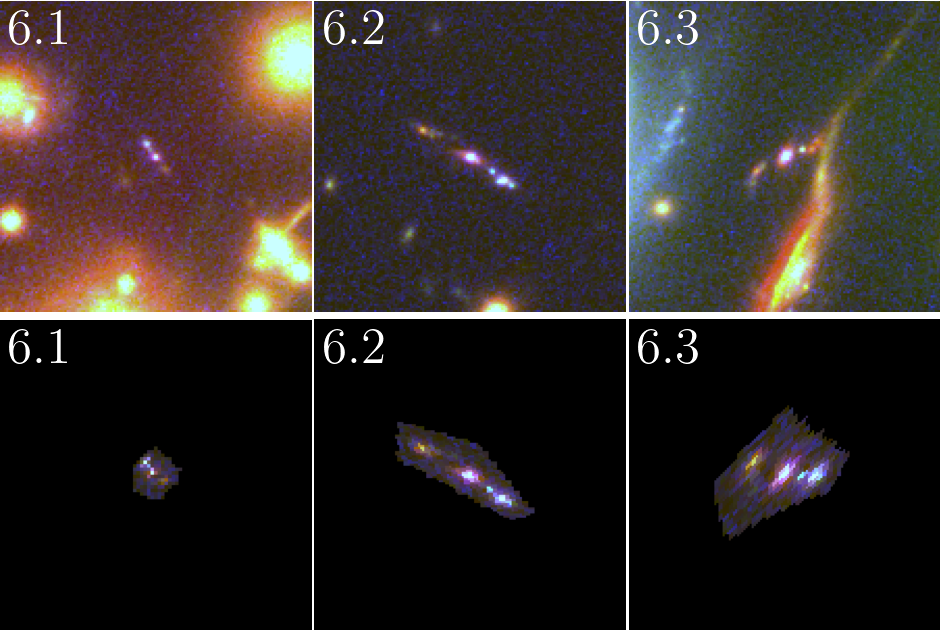}\label{subfig:Sys6}
    \end{minipage}
    \hfill
    \begin{minipage}[t]{0.492\textwidth}
    \vspace{0pt}
    \centering
    \includegraphics[width=\linewidth]{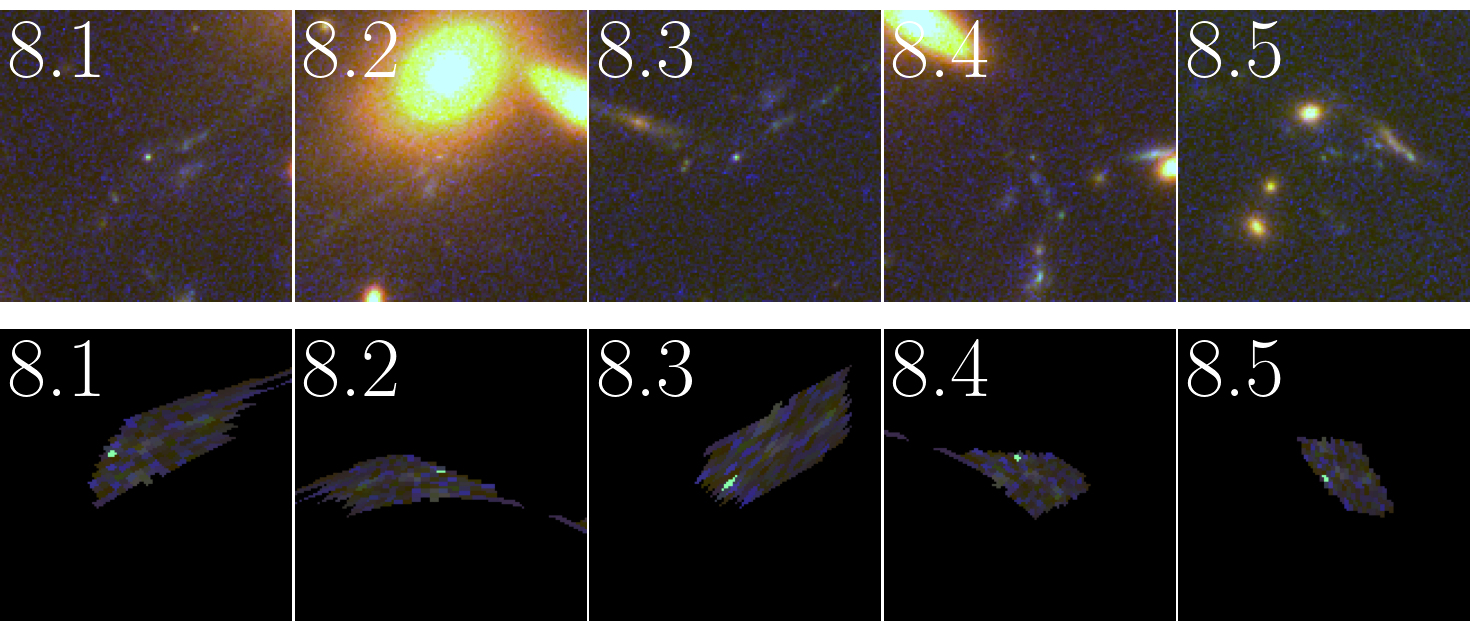}\label{subfig:Sys8}\\[2.5em]
    \includegraphics[width=\linewidth]{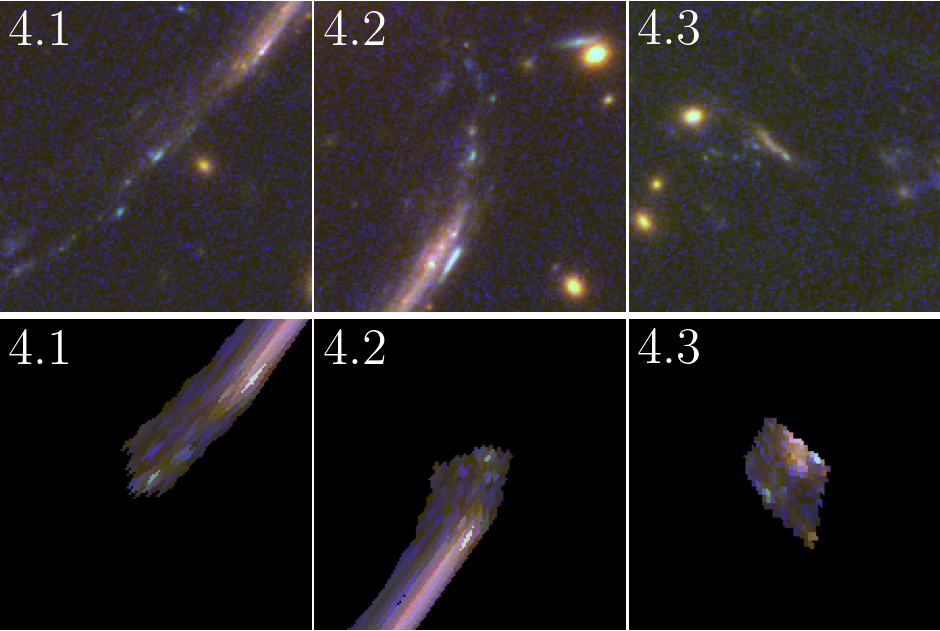}\label{subfig:Sys4}
    \end{minipage}\\[.5em]
    
    \includegraphics[width=\textwidth]{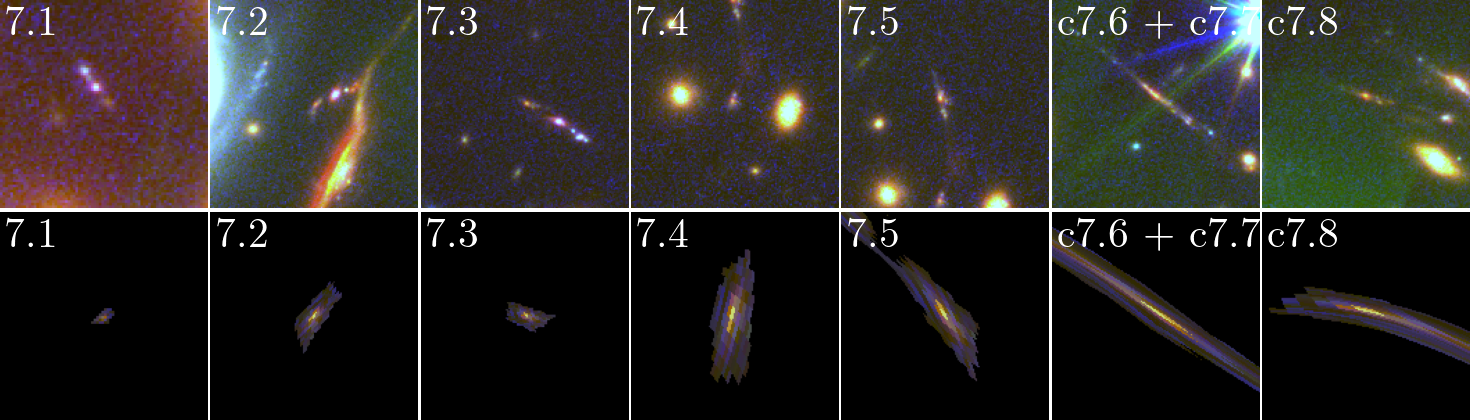}\label{subfig:Sys7}

    \caption{Reproduction of several systems by our model. For each system we delens one image -- typically the first one -- to the source plane and back to the image plane, displaying the reproduction of the other images in that system. 
    The upper row, for each system, shows the images as they appear in the RGB color image of the cluster (Figure\,\ref{fig:cc}), and the bottom row, their reproduction by our model. As can be seen, the model does a very good job in reproducing the appearance of multiple images, strengthening their identification and showcasing its robustness.}
    \label{fig:Sys_relensed}
\end{figure*}

\smallskip

\section{Lensing Analysis}\label{sec:analysis}
We use here the Light-Traces-Mass (LTM) modeling method developed by \citet[][see also \citealt{Broadhurst2005a}]{Zitrin2009_cl0024,Zitrin2014CLASH25}. 
This method has been extensively used on large samples of clusters, and thanks to the assumption that light traces mass, excels in matching up multiply imaged galaxies even before a model is fully minimized. 
It is thus a powerful method in particular for newly analyzed clusters. The multiple-image matching is performed by constructing an initially well-guessed model (using only the photometry of cluster members) and then delensing-relensing arclets to predict the observed appearance of counter images, which are then searched for in the observations.
The method is described in greater detail in \citet{Zitrin2014CLASH25} and we briefly summarize it here.

The model is composed of three components.
First, a map of cluster galaxies is constructed by assigning a power-law mass density profile to each galaxy, where the exponent is a free parameter of the model. The exponent is the same for all galaxies, and the mass (or weight) of each galaxy is scaled in proportion to its luminosity. The mass distribution of each galaxy is assumed to be circular, except the western BCG here, whose ellipticity is set to the measured values from SExtractor.
Then, a dark matter map is constructed by smoothing the resulting map of cluster galaxies, using a Gaussian kernel whose width is also a free parameter of the model. The relative weight between the galaxy and dark matter components, and the overall normalization, are free parameters in the model as well. Finally, an external shear is added, which allows for more freedom introducing effective ellipticity to the model. 
In addition, we also leave here the weight (i.e.\ mass normalization) of key bright galaxies free to be optimized by the model -- these are marked in Fig.\,\ref{fig:cc}. 
It is important to note that none of the lensed galaxies or multiple-image systems we find here has a spectroscopic redshift measurement. 
Since the exact redshifts of the lensed galaxies are unknown, we adopt, as an initial guide, their photometric redshift estimates but allow their redshift to be optimized by the model using a wide prior. In practice, this is done through a uniformly distributed prior range on the distance ratio, $D_{LS}/D_{S}$, typically spanning the redshift range from $z \sim 1.5 - 2$ to $z\gtrsim6$, where $D_{LS}$ and $D_{S}$ are the angular-diameter distances between the lens and source, and to the source, respectively.

Once multiple image systems are iteratively found using a set of preliminary lens models, as well as color, symmetry, appearance and photometric redshift information, a final model is constructed. The model is optimized using a MCMC, minimizing the $\chi^2$ function that quantifies the distance between the predicted positions of multiple images and their observed positions.

Because LTM needs to be anchored (i.e.\ scaled) to a reference redshift, we choose System\,3 (see Fig.\,\ref{fig:cc} and \ref{fig:Sys_relensed}) for its robust photometric redshift estimate and because it is well distributed across the cluster field. We anchor $z_3 = 3.5$, as estimated using the photometric redshift results (see Table\,\ref{tab:multTable}). To account for the uncertainty in the redshift of the system, we use two other identical setup models but anchored at $z_3 = 3.25$ and $3.75$ following the 1$\sigma$ photometric redshift uncertainty of $\Delta z = 0.25$. All uncertainties derived hereafter take this propagated uncertainty into account.

\begin{figure}
    \centering
    \includegraphics[width=\columnwidth]{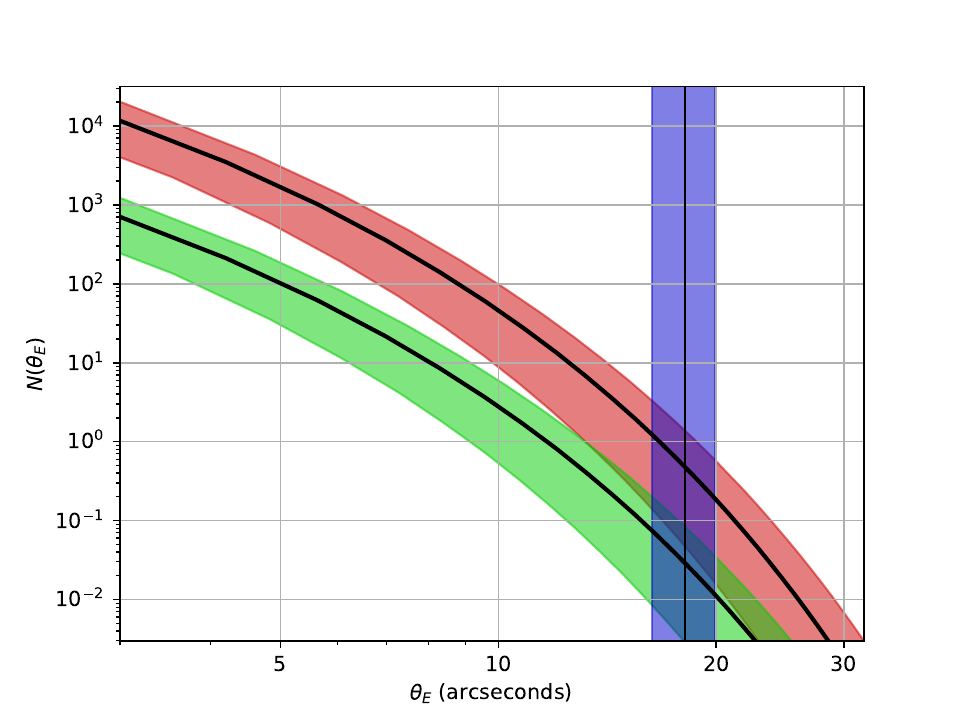}
    \caption{The expected distribution of Einstein radii around $z_{l}\sim1.07$ for a source at $z_{s}=3$, across the sky (shaded red), and for the SPT survey region (2,500 deg$^2$; shaded green), versus the Einstein radius of \clname\, $\theta_{E}=18.1\pm1.8\arcsec$ (vertical blue shaded region), for the same source redshift. The (nominally, 1$\sigma$) scatter in the Einstein radius distribution was computed by probing clusters in the range $z=1-1.2$, and from the scatter in the adopted $c-M$ relation (for more details see Section \ref{subsec:mass_discussion}). As can be seen, clusters with such Einstein radii are expected to be rare at this redshift (see discussion in text).}
    \label{fig:thetaEDist}
\end{figure}

\section{Results and Discussion} \label{sec:discussion}
\subsection{Mass model}

We mark the identified multiple-image systems, and the critical curves derived from the best-fit model on Figure\,\ref{fig:cc}.
In total, we detect 35 multiple images and a few additional candidates, belonging to at least 10 background sources (some systems, such as System\,6, contain clumps that may, in principle, be separate galaxies). 
All securely identified images of these 10 systems were used to constrain the lens model, at the exception of System\,7, which was optimized with System\,6 (see Section \ref{ref:lensing_features}); this amounts to 30 images used as constraints.
We also identify 6 other candidate systems, comprising 16 multiple image candidates, which can be confirmed with spectroscopic follow-up.

The large number of systems, and prominent lensing features seen in Fig.\,\ref{fig:cc}, are rather surprising for a $z_l > 1$ cluster and show its importance.
The model was minimized using 39 constraints and 28 free parameters, resulting in 11 degrees-of-freedom. The final model has a reduced $\chi^2\simeq68/11$ and a r.m.s.\ $\simeq1.2\arcsec$ in reproducing the multiple images. While slightly larger than typical high-end parametric lens models (e.g.\ \citealt{Richard2014FF, Kawamata2016, 2024MNRAS.533.2242F}), this is typical (and in fact quite low) for LTM models, especially for such complex clusters \citep[e.g.][]{Zitrin2017, Acebron2019RXC0032}.
Given it is coupled to the light distribution, the LTM methodology is less flexible on one hand, resulting in a somewhat higher r.m.s.\ as seen here, but on the other hand, allows for a strong, initial prediction power \citep{Carrasco2020,Zalesky2020MNRAS.498.1121Z}, particularly useful for complex clusters analyzed for the first time \citep[e.g.][]{Zitrin2017}. 
We make our resulting lens model publicly available\footnote{\dataset[doi: 10.5281/zenodo.16534337]{https://doi.org/10.5281/zenodo.16534337}}.

The mass density distribution in units of the critical density for lensing and the magnification maps are both shown in Fig.\,\ref{fig:kappamu}. 
The critical curves for sources at redshifts $z_s=2,3,$ and 9, enclose, respectively, areas of $\mathcal{A}_{z_s=2} = 0.09 \pm 0.02$ arcmin$^2$, $\mathcal{A}_{z_s=3} = 0.28 \pm 0.06$ arcmin$^2$ and $\mathcal{A}_{z_s=9} = 0.68 \pm 0.13$ arcmin$^2$. 
These correspond, respectively, to effective Einstein radii of $\theta_{E, z_s=2}=10.0\pm1.0\arcsec$, $\theta_{E, z_s=3}=18.1\pm1.8\arcsec$ and $\theta_{E, z_s=9}=27.9 \pm 2.8\arcsec$.
The $z_s = 2, 3$, and 9 critical curves enclose projected masses, respectively, of $M(\in \mathcal{A}_{z_s=2}) = (6.1\pm0.9)\times10^{13}\,M_{\odot}$, $M(\in \mathcal{A}_{z_s=3}) = (1.4\pm 0.2)\times10^{14}\,M_{\odot}$ and $M(\in \mathcal{A}_{z_s=9}) = (2.6\pm0.4)\times10^{14}\,M_{\odot}$.

The total projected (cylindrical) mass density enclosed within 200\,kpc from the cluster center, defined here between the two main mass clumps, i.e.\ (RA; Dec) = (86.6486467; -53.7599104) deg, is $M(<200\,\mathrm{kpc}) = (1.9 \pm 0.3) \times 10^{14}\,\Msol$, and within 500\,kpc, $M(<500\,\mathrm{kpc}) = (6.5 \pm 1.0) \times 10^{14}\,\Msol$.
This former mass is similar to that of typical massive lensing clusters usually observed at lower redshifts \citep[see e.g.][]{FoxMahlerSharon2022}, including the HFF sample for which $M(<200\,\mathrm{kpc})$ is in the range $(1.5 - 2.3) \times 10^{14}\,\Msol$, even though these are observed when the Universe is $\sim 3-4$\,Gyr older. 

In order to compare with previous mass measurements of this cluster by other probes such as from X-ray and SZE data, or from a previous HST weak-lensing analysis, we fit a spherical NFW profile \citep{Navarro1996} to the best-fit model density profile. We obtain an \emph{extrapolated} (3D) mass of $M_{500,c}^{\rm SL, NFW} = (7.2\pm 0.5) \times 10^{14}\,\Msol$\footnote{$M_{\Delta, c}$ is defined as the mass within a 3D sphere of radius $R_{\Delta}$, where the average density is $\Delta \cdot \rho_c (z)$, with $\rho_c (z)$ being the critical density of the Universe at redshift $z$, i.e.\ $R_{\Delta} = \left\{ R \mid \frac{M(< R)}{(4/3) \pi R^3} = \Delta \cdot \rho_c (z) \right\}$.}. The X-ray, SZE, and weak-lensing masses reported in the literature in a similar radius are
$M_{500}^{X} = (5.33 \pm 0.62) \times 10^{14}\,\Msol$ \citep{2011ApJ...738...48A},  $M_{500}^{\rm SZ} = (5.05 \pm 0.82) \times 10^{14}\,\Msol$ \citep{SPT2010ClustersV1}, and $M_{500}^{\rm WL} = (3.7^{+2.6}_{-2.3} \pm 0.8 \pm 0.5) \times 10^{14}\,\Msol$, \citep{Schrabback2018MNRAS.474.2635S}, respectively.
As this comparison is based on many assumptions (extrapolation, spherical symmetry, NFW profile, hydrostatic equilibrium of the intra-cluster medium), a variety of factors, such as projection effects or non-thermal pressure may explain this apparent few $\sigma$ scatter between the different measurements \citep{Meneghetti2010b, 2012NJPh...14e5018R, 2016arXiv160804388M}. 
For comparison, the hydrostatic-to-lensing mass ratio (or bias) we find for $M_{500}^{\rm HSE} / M_{500}^{\rm Lens} = 0.74$ is similar to the one found by \citet{2024A&A...682A.147M} for a sample of clusters at redshifts $0.05 - 1.07$.

\subsection{A few spectacular lensing features} \label{ref:lensing_features}
Some of the lensed sources show unique features. 
For instance, the prominent System\,8
forms what appears to be a hyperbolic-umbilic configuration \citep[a ring-shaped structure of four off-centered images, and typically an additional outer image, see][]{2009MNRAS.399....2O, 2023MNRAS.522.1091L, 2024OJAp....7E..91M}, including a giant arc designated here as System\,4 (see Fig.\,\ref{fig:cc} and Table\,\ref{tab:multTable}).
We note that Systems 4 and 8 may be two different parts of the same background galaxy. Similarly, System\,11 may in fact consist of counter images of System\,1.

Also of interest, System\,5 comprises an obvious nuclear point source, embedded in a very-dim arc, multiply imaged five or six times where three of the images wrap a cluster galaxy forming an Einstein quad.
The point-like appearance is reminiscent of other multiply imaged AGN detected in the same manner \citep[e.g.][]{Furtak2023AGN, 2024Natur.628...57F}. 
A zoom-in on the AGN candidate is displayed as an inset in Fig.\,\ref{fig:cc}.
Only a few galaxy cluster-lensed multiply imaged AGN are known to date \citep{2025ApJ...987..194C}.
If confirmed spectroscopically, this will supply another multiply imaged AGN with a range of time delays between its images that can then be used for measurements of the expansion rate of the Universe or for efficient reverberation mapping of super-massive black holes \citep[e.g.][]{Williams2021ApJ...915L...9W, 2023ApJ...959..134N,Golubchik2024ApJ...976..108G,Ji2025arXiv250113082J}.
Our lens model suggests that the Einstein cross images around the cluster member, i.e.\ images 5.3-5.6, arrive first within days or weeks of each other, and about $\sim$40-50 years before the other two images (5.1 and 5.2), which are predicted to arrive weeks from one another. We leave a more detailed investigation of the time delays to future works in case the AGN candidate is indeed confirmed to be a variable source.

Another system (System\,9), is a very red dusty object, at $z\sim$ 3, invisible in the optical HST data. 
System\,7 is also remarkable, as some of its multiple images are lensed into a complex \emph{giant} arc with very high magnifications, potentially allowing uniquely zoomed-in view into the various stellar clumps seen therein (similar to other lensed clumps seen uniquely with JWST, e.g.\ \citealt{2023ApJ...945...53V, 2025A&A...693A..33L}). Given the complexity of the arc and the fact that it is much more magnified than the other counter images of that system, it is hard to determine which substructures are counter-imaged, and thus we do not use this system as constraints, and leave a more detailed investigation of this arc to future work. 
Nonetheless, as displayed in Fig.\,\ref{fig:Sys_relensed}, our lens model naturally reproduces the arc, acting as another evidence for its robustness.
More details on the specific images are given in Table \ref{tab:multTable}.

\subsection{An uncommonly massive galaxy cluster?} \label{subsec:mass_discussion}
To get a sense of the rarity of the strong-lensing properties of the cluster, we compare its Einstein radius to a semi-analytic calculation using a cosmological halo mass function. We adopt the \citet{Tinker2008MF} mass function, and examine the expected number of lenses per given Einstein radius, in the redshift range $z_l=1 - 1.2$. 
To do this, we run over the mass function in $10^{14}\,M_{\odot}$ bins from $10^{14}\,M_{\odot}$ to $3\times10^{15}\,M_{\odot}$, and over the above redshift range in $\delta z=0.1$ bins. For each redshift and for each mass bin, we calculate the distribution of Einstein radii \citep{SadehRephaeli2008} by assuming a NFW halo \citep{Navarro1996} for the clusters and adopting the concentration-mass relation and scatter from \citet{Meneghetti2014CLASHsim}.
The distribution of Einstein radii $N(\theta_E)$ is then obtained by integrating over the comoving volume per each redshift bin and over all mass bins (for more details, see also \citealt{Golubchik2024ApJ...976..108G}).   
The distribution is shown in Fig.\,\ref{fig:thetaEDist}, where we show both the expected all-sky distribution and the same distribution normalized to the SPT survey area -- with which \clname\ was first detected.
We find that, at $z_{l}\simeq1.07$, the largest observed Einstein radius across the sky should typically be of the same order as observed for \clname, but the chances to find such a cluster in the SPT survey are much smaller, given the small fraction of the SPT survey area compared to the whole sky, $\sim$0.06. This somewhat simplistic treatment has some obvious caveats. 
For instance, the adopted mass function is simulation based, and in addition, the small simulation volume is likely missing the largest clusters \citep{2014MNRAS.437.3776W}, and thus the corresponding number density of large Einstein radii based on the mass function we adopt should be considered a lower limit. Moreover, our calculation does not take into account the lens ellipticity or merging state, and the $c - M$ relation may differ at higher redshifts. Nonetheless, the calculation generally shows that, indeed, such lensing configurations should be rare at $z\gtrsim1$.

Nevertheless, several other distant lensing galaxy clusters that were analyzed in recent years show prominent lensing features, and correspondingly, large central projected masses. For example, using similar data from the SLICE JWST survey, \citet{Cerny25} modeled the galaxy clusters SPT-CL J0516-5755 ($z_l=0.966$) and SPT-CL J2011-5228 ($z_l = 1.064$). These two clusters have Einstein radii of $\theta_{E, z_s=2} = 6-7''$ for a source at $z_s = 2$, and $\theta_{E, z_s = 9} = 22-33''$ for a source at $z = 9$. For comparison, our present analysis of \clname\ yields Einstein radii of $\theta_{E, z_s=2} = 10.0 \pm 1.0''$ and $\theta_{E, z_s=9} = 27.9\pm2.8''$ respectively, i.e.\ displaying similar or stronger lensing properties.
The enclosed 2D projected mass within 500\,kpc found by \cite{Cerny25} for these two clusters is $M(<\mathrm{500\,kpc}) = 5.10^{+0.42}_{-0.18} \times 10^{14}\,\Msol$ and $7.73^{+1.75}_{-0.00} \times 10^{14}\,\Msol$ respectively, i.e.\ similar to the $M(<\mathrm{500\,kpc}) = (6.5\pm1.0) \times 10^{14}\,\Msol$ we find here for \clname.

Other examples of prominent lensing clusters around $z_l\sim1$ include those previously analyzed with HST data, such as MCXC\,J1226.9+3332 \citep[$z_l = 0.89$, $\theta_{E, z_s = 2} = 14.5\pm1.5$\arcsec,][]{Zitrin2014CLASH25},
SPT-CL\,J0615-5746 \citep[$z_l=0.97$, $M(<200\,\mathrm{kpc}) = 2.51^{+0.15}_{-0.09} \times 10^{14}\,\Msol$,][]{2018ApJ...863..154P}, SPT-CL\,J0356-5337 \citep[$z_l = 1.04$, $\theta_{E, z_s = 3} \simeq 14 \arcsec$,][]{2020ApJ...894..150M, smith2025z103mergingclustersptcl}, and El Gordo \citep[ACT-CL\,J0102-4915, $z_l = 0.87$, $M(<500\,\mathrm{kpc}) = (8.3 \pm 0.3) \times 10^{14}\,\Msol$,][]{Diego2022Quyllur}.
These clusters exhibit similar strong-lensing properties or central mass as \clname, and hint that \clname\ analyzed in this work, although expected to be rare for its redshift, is in fact quite common. Thanks to HST and JWST data in particular, it is now becoming evident that galaxy clusters at $z_l \gtrsim 1$ can be sufficiently massive and concentrated to form prominent strong lenses, especially for higher redshift sources. 

Over a 100 clusters were detected at redshifts of $z\sim1$ or above with the SPT \citep[e.g.][]{BleemSPTCat2015ApJS..216...27B}.
Of these, about a couple dozen are included in the SLICE program, which is designed to study the cosmological evolution of the high-mass end of the halo mass function using strong lensing, over a redshift range from $z \sim 1.9$ to $z \sim 0.2$ -- corresponding to about 8 Gyr of cosmic time. 
It will thus allow to investigate whether these other high-redshift clusters exhibit similarly impressive lensing features as seen in \clname, and how these evolve with cluster redshift.

\section{Summary} \label{sec:conclusion}
In this work, we present a first strong-lensing model for the $z=1.067$ galaxy cluster \clname, enabled thanks to new JWST/NIRCam observations from the SLICE survey, complemented by previous HST data. The cluster reveals as a very prominent gravitational lens. Using the LTM mass-modeling technique coupled with color information, we identify 16 sets (and candidates) of multiply imaged galaxies spanning, according to their photometry and the best-fit mass model, the redshift range $\sim2-6$. 
The multiply imaged systems include various interesting features, such as a candidate, sextuply lensed point-source which may be an AGN; various dusty galaxies at cosmic noon; an extremely magnified stretched arc; and a hyperbolic-umbilic configuration. 

The conspicuous lensing features align well with the significant Einstein radius of the cluster, $\theta_{E, z_s=3}=18.1\pm1.8\arcsec$ for $z_{s}=3$, or $\theta_{E, z_s=9}=27.9\pm2.8\arcsec$ for $z_{s}=9$, and with its central projected mass density, $M(<200\,\mathrm{kpc}) = (1.9 \pm 0.3) \times 10^{14}\,\Msol$.
These values are similar to some of the best-studied lensing clusters such as those previously studied (e.g.\ in the frameworks of the CLASH, Hubble Frontier Fields or RELICS programs), even though the cluster is at a significantly higher redshift and thus seen at a significantly earlier cosmic time, where such prominent strong-lensing clusters are expected to be scarce. We compare this Einstein radius with the distribution of Einstein radii predicted for the cluster's redshift, and find that it is indeed expected to be rare, in particular when taking into account the SPT survey area. 
According to our simplified estimation presented in Section\,\ref{subsec:mass_discussion}, only a few clusters are expected to show equal or larger Einstein radii, at $z\simeq1-1.2$, across the whole sky. Nonetheless, several other prominent lenses have been discovered around redshift $z \sim1$ in recent years, including those analyzed by \citet{Cerny25} using similar JWST data from the SLICE survey.

This work thus highlights the advantage of using JWST, thanks to its sensitivity and wavelength coverage, in cluster evolution and lensing-related studies, and their extension to higher lens redshifts of $z\gtrsim1$. Analysis of more $z > 1$ clusters with JWST in the SLICE program is forthcoming.

\begin{acknowledgments}
The authors thank the reviewer of this work for useful comments, which helped to improve the article.
The authors would like to thank G. Brammer and the DAWN JWST Archive.
AZ acknowledges support by grant 2020750 from the United States-Israel Binational Science Foundation (BSF) and grant 2109066 from the United States National Science Foundation (NSF), and by the Israel Science Foundation Grant No. 864/23. 
MJ is supported by the United Kingdom Research and Innovation (UKRI) Future Leaders Fellowship `Using Cosmic Beasts to uncover the Nature of Dark Matter' (grant number MR/X006069/1). ML acknowledges CNRS and CNES for support.

This work is based on observations made with the NASA/ESA/CSA \textit{James Webb Space Telescope} (JWST) and with the NASA/ESA \textit{Hubble Space Telescope} (HST). The data were obtained from the \texttt{Barbara A. Mikulski Archive for Space Telescopes} (\texttt{MAST}) at the \textit{Space Telescope Science Institute} (STScI), which is operated by the Association of Universities for Research in Astronomy (AURA), Inc., under NASA contract NAS 5-03127 for JWST. These observations are associated with the JWST GO program number 5594 and HST 12477, 15294.
The HST specific observations analyzed can be accessed via \dataset[doi: 10.17909/284j-8g39]{https://doi.org/10.17909/284j-8g39}, while the JWST data set is at \dataset[doi: 10.17909/t661-kf88]{https://doi.org/10.17909/t661-kf88}.
The maps associated to the lens model, as well catalogs cited in this article may be downloaded from the Zenodo dataset \dataset[doi: 10.5281/zenodo.16534337]{https://doi.org/10.5281/zenodo.16534337}.
\end{acknowledgments}

\tabletypesize{\scriptsize}
\begin{deluxetable*}{|l|c|c|c|c|c|}
\tablecaption{\small{Multiple Images and Candidates} \label{tab:multTable}}
\tablehead{
\colhead{Arc ID
} &
\colhead{R.A.
} &
\colhead{Dec.
} &
\colhead{$z_{\rm phot}~50\%$ [16\%--84\%]
} &
\colhead{$z_{\rm model}$ [16\%--84\%]
} &
\colhead{Comments}
}
\startdata
1.1 & 05:46:37.1684 & -53:45:29.876 &  --- & 2.76 [2.58--2.79] & Relensed images predicted in the west, \\
1.2 & 05:46:36.9976 & -53:45:30.179 & ---  & " & maybe corresponding to System\,11\\
\hline 
2.1 & 05:46:36.9549 & -53:45:27.906   & 1.58 [1.33--1.95]  &  1.73 [1.72--1.94]  &  \\
2.2 & 05:46:37.2526 & -53:45:28.578   & ---  & "  &  \\
c2.3 & 05:46:38.2754 & -53:45:49.833   & 1.65  [1.35--2.07]  &  " & The relensed image corresponds with c2.3 \\
\hline
3.1 & 05:46:39.0491 & -53:45:27.604   & 3.47 [3.21--3.78] &  3.50 [3.25--3.75]$^{\star}$ &\\
3.2 & 05:46:35.6897 & -53:45:35.207   & 3.51 [3.23--3.85] &  " &\\
3.3 & 05:46:37.5895 & -53:45:55.317   & 3.47 [3.23--3.75] &  " &\\
3.4 & 05:46:36.5909 & -53:45:08.624   & 3.48 [3.25--3.75] &  " &\\
3.5 & 05:46:36.4313 & -53:45:09.137   & --- & " &\\
\hline
4.1 & 05:46:33.5957 & -53:45:51.184   & 1.85 [1.40--3.34] & 3.76 [3.31--3.87]  & \\
4.2 & 05:46:33.1748 & -53:45:45.583    & 1.72 [1.37--2.91] &  " & Possible ALMA continuum detection\\
4.3 & 05:46:33.9035 & -53:45:19.681   & 1.86 [1.43--2.64] &  " & \citep{2018ApJ...853..195W} \\
\hline
c4.11 & 05:46:33.3698 & -53:45:49.174& 2.31  [1.46--3.37] & --  &\\
c4.21 & 05:46:33.2156 & -53:45:47.095 & " &  --   &\\
\hline
5.1 & 05:46:35.5420 & -53:45:45.915   & 3.29 [2.16--3.58] &  3.41 [3.22--3.44]  & AGN candidate\\
5.2 & 05:46:35.7182 & -53:45:48.462   & 3.25 [2.58--3.58] &  "  & \\
5.3 & 05:46:34.8683 & -53:45:13.497  & 3.14 [2.16--3.58] &  "  &\\
5.4 & 05:46:34.8157 & -53:45:13.621   & 3.01 [1.97--3.44] &  "   &\\
5.5 & 05:46:34.7178 & -53:45:14.797   & 2.95 [2.16--3.45] &  "   &\\
c5.6 & 05:46:34.8619 & -53:45:14.292 & --- &  "  & Fainter, candidate image\\
\hline
6.1 & 05:46:37.5542 & -53:45:27.352   & 5.03 [4.78--5.25]  &  6.94 [6.21--7.06]  &\\
6.2 & 05:46:36.9310 & -53:46:10.274   & 5.01 [4.92--5.10]  &  "  &\\
6.3 & 05:46:39.1766 & -53:45:16.814   & 4.90 [4.79-5.03] &  "   &\\
\hline
7.1 & 05:46:37.5350 & -53:45:27.571   & --- & 6.94 [6.21--7.06]  & Optimized with System\,6\\
7.2 & 05:46:37.0333 & -53:46:09.754   & 5.65 [1.61--6.19] &  "   &\\
7.3 & 05:46:39.2391 & -53:45:17.106   & 3.59 [2.95--4.11] &  "   &\\
7.4 & 05:46:34.8151 & -53:45:25.998   & 5.03 [4.30--5.23] &  "   &\\
7.5 & 05:46:34.8151 & -53:45:23.047   &  5.43 [5.15--5.66] &  "  & \\
c7.6 & 05:46:35.4921 & -53:45:16.163   & --- &  "  & Images c7.6--c7.8 form a very long arc\\
c7.7 & 05:46:35.8078 & -53:45:14.014   & 5.44 [1.35--5.99] &   "  &which consists likely of counter images\\
c7.8 & 05:46:36.8846 & -53:45:08.964    &  6.07 [1.77--6.99] &  "   &of systems 6+7; exact configuration unclear\\
\hline
8.1 & 05:46:34.1448 & -53:45:44.274   & 2.25 [1.63--3.32] & 3.99 [3.51--4.00] & Hyperbolic-umbilic-like system;\\
8.2 & 05:46:33.7016 & -53:45:41.942   & --- &   " & system may be related to System\,4\\
8.3 & 05:46:33.9781 & -53:45:53.970    & 2.68 [1.69--3.43] & " &\\
8.4 & 05:46:33.1879 & -53:45:43.434   & --- & "  &\\
8.5 & 05:46:34.0748 & -53:45:19.680   & --- &   "  &\\
\hline
9.1 & 05:46:39.1740 & -53:45:19.026   & 3.58 [3.31--3.84] & 2.88 [2.88--3.01]  & A known, dusty, star-forming galaxy;\\
9.2 & 05:46:39.2523 & -53:45:20.672  & --- & " & ALMA detected \\
9.3 & 05:46:39.2773 & -53:45:42.406   &  3.52 [2.01--4.12] &  "  & \citep{2018ApJ...853..195W}\\ 
\hline
10.1 & 05:46:39.9678 & -53:45:35.575   & --- & 3.70 [3.53--3.74]  &\\
10.2 & 05:46:39.9047 & -53:45:38.160   & --- &  "   &\\
\hline
c11.1 & 05:46:37.8396 & -53:45:13.256   & 1.76 [1.37--2.58] & ---   &Images of this system may be \\
c11.2 & 05:46:38.9357 & -53:45:25.280   & 1.78 [1.44--2.56] &  ---   &counter images of system 1\\
c11.3 & 05:46:38.4399 & -53:45:46.624   & 1.80 [1.43--2.52] &  --- &\\
\hline
c12.1 & 05:46:38.1034 & -53:45:17.095   & 2.15 [1.59--2.89] &   --- &  $z_{\rm phot}$ of 12.1 was measured on the arc,\\
c12.2 & 05:46:38.3263 & -53:45:18.535   & 2.14 [1.86--2.48] &  ---  & between the two clumps 12.1 and 12.2\\
\hline
c13.1 & 05:46:41.2089 & -53:45:21.101   & 2.92  [1.85--3.15] &  --- & Galaxy-galaxy strong lensing\\
c13.2 & 05:46:41.1951 & -53:45:19.834   & ---  &   --- & $z_{\rm phot}$ for 13.1 was measured\\
c13.3 & 05:46:41.1794 & -53:45:19.532   & --- &   --- & on the nearby nucleus\\
\hline
c14.1 & 05:46:38.5037 & -53:45:11.698  & --- &   --- & \\
c14.2 & 05:46:38.7812 & -53:45:12.904   & --- &   --- &\\
c14.3 & 05:46:38.9436 & -53:45:14.746   & --- &   --- &\\
\hline
c15.1 & 05:46:34.9388 & -53:45:39.220   & --- &   --- &  \\
c15.2 & 05:46:35.9804 & -53:45:55.658   & 1.83 [1.40--3.48] &  --- &\\
c15.3 & 05:46:35.0187 & -53:45:13.800   & 1.68 [1.32--3.18] &   --- &\\
\hline
\enddata
\tablecomments{$\emph{Column 1:}$ arc ID. ``c'' stands for candidate where identification was more ambiguous, or if image was not used as constraint.\\
$\emph{Columns 2 \& 3:}$ RA and DEC in J2000.0.\\
$\emph{Column 4:}$ Photometric redshift 50\% [16\%--84\%] percentiles from \texttt{Bagpipes}.\\
$\emph{Column 5:}$ Predicted and 64\% C.I. redshift by our LTM lens model, for systems whose redshift was left to be optimized in the minimization.\\
$\emph{Column 6:}$ Comments.\\
For systems 12 and 13, some images were not identified in the photometric catalog, and thus we quote a photometric-redshift for a nearby clump or region from the same arc, as also mentioned in the relevant comments column.\\
$\star$ All the uncertainties on model redshifts, and across this article, take into account the 1$\sigma$ redshift uncertainty for System\,3 ($z_s = 3.5 \pm 0.25$) by running three different mass models, anchored at $z_s = 3.25$, 3.5 and 3.75.
}
\end{deluxetable*}

%%%%%%%%%%%%%%%%%%%%%%%%%%%%%%%%%%%%%%%%%%%%%%%%%%%%

\bibliographystyle{aasjournalv7}
\bibliography{MyBiblio}
\end{document}